\documentclass[epj]{webofc}
\usepackage[varg]{txfonts}   
\wocname{EPJ Web of Conferences}
\woctitle{INPC 2013}

\usepackage{bm}

\begin{document}
\title{A microscopic model beyond mean-field:}
%
%
\subtitle{from giant resonances properties to the fit of new effective interactions}

\author{M. Brenna\inst{1,2}\fnsep\thanks{\email{marco.brenna@mi.infn.it}} \and
        G. Col\`o\inst{1,2} \and X. Roca-Maza\inst{1,2} \and P. F. Bortingon\inst{1,2}
        \and K. Moghrabi\inst{3} \and M. Grasso\inst{3}
}

\institute{Dipartimento di Fisica, Universit\`a degli Studi di Milano, Via Celoria
16, I-20133, Milano, Italy
\and
INFN, Sezione di Milano, Via Celoria 16, I-20133, Milano, Italy
\and
Institut de Physique Nucl\'eaire, Universit\'e Paris-Sud, IN2P3-CNRS, F-91406 Orsay Cedex, France
}

\abstract{%
A completely microscopic beyond mean-field approach has been elaborated to overcome some intrinsic limitations of
self-consistent mean-field schemes applied to nuclear systems, such as the incapability to produce some properties of
single-particle states (e.g. spectroscopic factors), as well as of collective states (e.g. their damping width and their
gamma decay to the ground state or to low lying states).
Since commonly used effective interactions are fitted at the mean-field level, one should aim at refitting them including the
desired beyond mean-field contributions in the refitting procedure. If zero-range interactions are used, divergences arise.
We present some steps towards the refitting of Skyrme interactions, for its application in finite nuclei.
}
\maketitle
\section{Introduction}
\label{intro}
Self-consistent mean-field (SCMF) models have become increasingly sophisticated and reliable in the overall description of
bulk nuclear properties (e.g. masses, radii and deformations), as well as of collective excitation, like
giant resonances (GR) \cite{Bender2003}. These are described as a coherent superposition of one particle-one hole
excitations, and are located in the energy range between 10 and 30 MeV. Thus, they can decay through the emission of
particles: this process is associated with the so-called escape width $\Gamma^{\uparrow}$. However, the most probable damping
mechanism is the coupling to progressively more complicated states (of 2p-2h, 3p-3h, ..., np-nh character). The associated
contribution to the total width, called spreading width ($\Gamma^{\downarrow}$), is the dominant one. Eventually, the
$\gamma$-decay width ($\Gamma_{\gamma}$), given by the coupling to the electromagnetic
field is a small fraction ($\le 10^{-2}$) of the total width. Despite this, the study of the $\gamma$ decay of GRs
has been considered a valuable tool for about 30 years \cite{Beene1989,Beene1990}.

Nevertheless, the SCMF approachs present well-known limitations. For instance, it does not reproduce, as a rule, the level
density around the Fermi energy. Moreover, the fragmentation of single particle states and the GR spreading widths are
outside the framework of these models.

One way to overcome these issues is the introduction of an interplay between single particle and collective degrees of
freedom, like vibrations. The basic ideas of the so-called particle-vibration coupling (PVC) models have been discussed in
Ref.~\cite{BMII}. Including these couplings, the standard shell model acquires a dynamical content: the average potential
becomes non-local in time or, which is the same, energy dependent~\cite{Mahaux1985}. Recently, some of us has developed a
fully microscopic
self-consistent model, based on Skyrme functionals, to treat properly single particle states~\cite{Colo2010}. In this
contribution we report on the application of this model to the calculation of inclusive (strength function) as well as
exclusive ($\gamma$-decay width) GRs properties.

If theories beyond mean-field are necessary, a problem of overcounting may arise since, so far, effective forces have been 
fitted at mean-field, thus including in an effective way a class of higher order correlations.
Therefore, we should expect to be obliged to re-fit the interactions at the required level of approximation. 
Moreover, if zero-range forces are used, like the Skyrme one, divergences arise when they are employed at beyond 
mean-field level. In the second part of this contribution, we present some preliminary results on the way to reabsorb 
these divergences in a simplified Skyrme interaction applied to finite nuclei.

\section{The giant resonances widths within the PVC approach}
\label{sec-1}
In this section we present the results obtained for two observables: the line shape of isovector giant quadrupole resonance
(IVGQR) in ${}^{208}$Pb,
and the width associated with the $\gamma$ decay of the isoscalar giant quadrupole resonance (ISGQR) in ${}^{208}$Pb 
to the ground state and to the first collective octupole state. Details about the formalism used and a more accurate
discussion of the results can be found in \cite{Roca-Maza2013} and \cite{Brenna2012}.

\subsection{The strength function}
\label{sec-2}

\begin{figure}
\centering
\sidecaption
\includegraphics[width=0.5\columnwidth,clip]{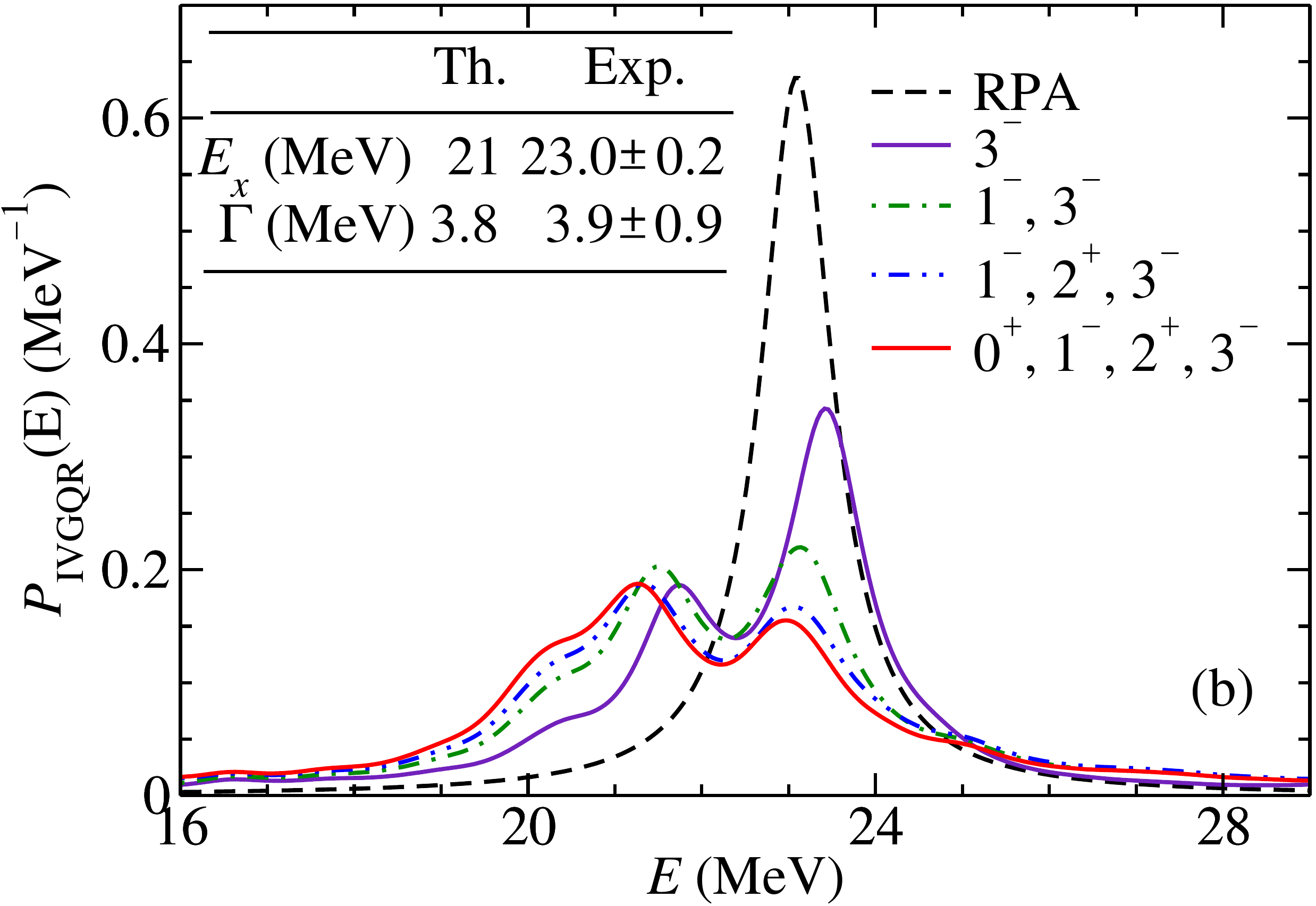}
\caption{(Color online) Probability P to find the IVGQR state at an energy E. Different curves are obtained when the phonons 
listed in the legend are used as intermediate states. 
The label RPA [black-dashed line] refers to the curve calculated in the RPA with a Lorentzian width of 1 MeV.
\label{fig-1}}
\end{figure}

The calculation that we report on here follows closely Ref.~\cite{Bortignon1981405}. In that paper, calculations of the GRs 
strength functions were based on the use of a phenomenological separable force in the surface coupling model.
The main novelty of this work is the consistent use of the Skyrme force in both HF+RPA diagonalization and in the PVC
vertex, added on top of the former.
In Fig.~\ref{fig-1} the probability per unit energy of finding the isovector quadrupole resonance state is plotted, 
calculated adopting the recent SAMi~\cite{Roca-Maza2012b} parametrization of the Skyrme force.
Phonons with multipolarity $L = 0,1,2,3$ and natural parity are included in the model space. 
The RPA peak, after the introduction of beyond mean-field correlations, is divided
into two parts. Although it decreases in magnitude, the centroid of the higher energy one is basically not affected by the
introduction of intermediate states of different multipolarity. On the other hand, the other peak broadens and is shifted to
lower energies. Eventually, the energy centroid can be set at 21 MeV and the spreading width is about 3.8 MeV, in good
agreement with the experimental findings (see inset in Fig.~\ref{fig-1}).

\subsection{The $\gamma$-decay width}
\label{sec-3}

\begin{table}
\centering
\caption{$\gamma$-decay widths $\Gamma_{\gamma}$. For details see the text and Ref.~\cite{Brenna2012}.\label{1-tab}}
\begin{tabular}{lcccc}
\hline
Interaction & $\mathrm{E}_{\mathrm{ISGQR}}$ [MeV] &
 $\Gamma_{\gamma}^{\mathrm{GS}}$ (eV) &
 $\mathrm{E}_{3^-}$ [MeV] &
 $\Gamma_{\gamma}^{3^-}$ (eV) \\
\hline
SLy5 \cite{Chabanat1998231} & 12.28 & 231.54 & 3.62 &  3.39 \\
SGII \cite{VanGiai1981} & 11.72 & 163.22 & 3.14 & 29.18 \\
SkP  \cite{Dobaczewski1984103} & 10.28 & 119.18 & 3.29 &  8.34 \\
LNS  \cite{Cao2006}  & 12.10 & 176.57 & 3.19 & 39.87 \\
\hline
Ref.~\cite{Beene1985} & 11.20 & 175 & -- & -- \\
Ref.~\cite{Bortignon198420} & 11.20 & 142 & 2.61 & 3.5 \\
Ref.~\cite{Speth1985} & 10.60 & 112 & 2.61 & 4 \\
\hline
Ref.~\cite{Beene1989} & 10.60 & 130 $\pm$ 40 & 2.61 & 5$\pm$5 \\
\hline
\end{tabular}
\end{table}

In the model described in Ref.~\cite{Brenna2012}, the $\gamma$ decay of the GRs to the ground state is evaluated at the 
RPA level, while the decay to low-lying collective states is accounted for at the first contributing order beyond the 
mean-field. It should be noted that only the direct decay can be computed in this model~\cite{Beene1985}.
In this contribution we focus on the ISGQR in ${}^{208}$Pb and its decay to the ground state and to the first $3^-$ state,
using different Skyrme interactions.
In Tab.~\ref{1-tab}, the obtained decay widths ($\Gamma_{\gamma}$) are listed, together with some results present in the
literature and with the experimental outcome. 

For the decay to the ground state the main issue is the overestimation of the energy of the resonance: the decay width is
proportional to the energy raised to $2\lambda+1$, being $\lambda$ the multipolarity of the transition.
If we rescale the energy to the experimental value, all the interactions can reproduce the experimental decay width within
the experimental error. On the other hand, concerning the decay to the $3^-$ state, only two interactions are in agreement
with the experiment. Actually, it should be recognized that it is just remarkable that Skyrme interactions can reproduce
the order of magnitude (few eV) of this exclusive observable, provided with the fact that this functional form are fitted to
reproduce basically macroscopic properties of nuclei at the scale of hundreds of keV.

\section{The problem of the divergences beyond mean-field}
As recalled in the Introduction, several beyond the mean-field quantities diverge if zero-range interactions are employed.
Until now,
the problem is circumvented by imposing a truncation to the model space with a somewhat arbitrary recipe, and this is clearly
unsatisfactory.
In Refs.~\cite{Moghrabi2010,Moghrabi2012}, the problem of the renormalization of the whole Skyrme interaction was faced in 
an infinite system with different degrees of neutron-proton asymmetry (from uniform to pure neutron matter): a new parameter,
namely the maximum value of the transferred momentum, was introduced among the Skyrme forces parameters.
In a finite system, like the nucleus, because of the absence of translational invariance, the transferred momentum is not
defined, opposite to what happens in the infinite system. However, we can write the velocity-independent part of the Skyrme
interaction as 
\begin{equation*}
   V(\bm{r}'_1,\bm{r}'_2,\bm{r}_1,\bm{r}_2) =
   \frac{1}{2\sqrt{2}} g \biggl( \frac{\bm{R}}{\sqrt{2}} \biggr) \delta_3(\bm{r}') \delta_3(\bm{r})
   \delta_3(\bm{R}-\bm{R}')=
   \frac{1}{2\sqrt{2}} g \biggl( \frac{\bm{R}}{\sqrt{2}} \biggr) v(\bm{r}',\bm{r}) \delta_3 (\bm{R}-\bm{R}'),
\end{equation*}
where $\bm{r}^{(\prime)}=\frac{\bm{r}^{(\prime)}_1-\bm{r}^{(\prime)}_2}{\sqrt{2}}$, $\bm{R}^{(\prime)} =
   \frac{\bm{r}^{(\prime)}_1+\bm{r}^{(\prime)}_2}{\sqrt{2}}$ and
$g(\bm{R})=t_0+\frac{t_3}{6}\left[\rho(\bm{R})\right]^{\alpha}$.
   We introduce a cutoff on relative momenta \cite{carl12}, modifying the term $v(\bm{r}',\bm{r})$, by including a cutoff
   $\lambda$ and $\lambda'$ (on particle 1 and 2),
\begin{equation*}
    v^{\lambda \lambda'}(\bm{r}',\bm{r}) = \frac{1}{\Omega} \int \mathrm{d}_3 k \mathrm{d}_3 k' \,
                                           \mathrm{e}^{i \bm{k}' \! \cdot \bm{r}'}
                                           v\left(\bm{k}', \bm{k}\right)
                                           \theta(\lambda - k) \theta(\lambda' - k')
                                           \mathrm{e}^{-i \bm{k} \! \cdot \bm{r}}
                                         = \frac{1}{4 \pi^4} \frac{\lambda^2 \lambda'^2}{r r'}
                                            j_{1}(r \lambda) j_{1}(r' \lambda')
\end{equation*}
and we use this new truncated interaction $v^{\lambda \lambda'}(\bm{r}',\bm{r})$ to compute the matrix elements relevant for 
the total energy of the system or the single particle energy, both computed at the second order on the many body theory.
The analysis of the results are currently undergoing.

\section{Conclusion}
We have seen that beyond mean-field correlations are really important to overcome some intrinsic limitations of the SCMF 
approach in nuclei. A completely microscopic model, based on the idea that single particle states can couple to collective
degrees of freedom (the so-called PVC), was recently developed. This model has been so far applied to single particle 
observables, like the self-energy, and, as reported here, to inclusive (strength and energy) and exclusive
($\gamma$ decay) properties of the GRs. The results obtained for the strength function of the IVGQR and the $\gamma$ decay 
of the ISGQR in ${}^{208}$Pb are in fairly good agreement with the experimental findings.

Anyhow, the employment of interactions fitted at mean-field level in a higher order framework is the main limitation of
these works. Moreover, zero-range forces cause the divergence of beyond mean-field quantities.
We presented here a first attempt to treat this problem in finite systems.

\begin{acknowledgement}
M.B. acknowledge the warm hospitality of the IPN Orsay, where part of this work was carried out.
\end{acknowledgement}

\bibliography{bibliography}

\end{document}